\documentclass{emulateapj}

\usepackage{ulem}  
\usepackage{color} 
\newcommand{\rel}{relativistic} 
\newcommand{\nonrel}{non-relativistic} 
\newcommand{\NL}{nonlinear} 
\newcommand{\TP}{test-particle} 
\newcommand{\synch}{synchrotron} 
\newcommand{\CD}{contact discontinuity} 
\newcommand{\tSNR}{t_\mathrm{SNR}} 
\newcommand{\EnSN}{E_\mathrm{SN}} 
\newcommand{\Mej}{M_\mathrm{ej}} 
\newcommand{\Inj}{\chi_\mathrm{inj}} 
\newcommand{\RadFS}{R_\mathrm{FS}} 
\newcommand{\RoRfs}{R/R_\mathrm{FS}}

\newcommand{\gameff}{\gamma_\mathrm{eff}}
\newcommand{\gamsk}{\gamma_\mathrm{sk}}
\newcommand{\effDSA}{\epsilon_\mathrm{DSA}}
\newcommand{\pcc}{cm$^{-3}$}
\newcommand{\np}{n_{p,0}}

\newcommand{\OxySix}{O$^{6+}$}
\newcommand{\OxySeven}{O$^{7+}$}
\newcommand{\OxyEight}{O$^{8+}$}

\newcommand{\SiTwelve}{Si$^{12+}$}
\newcommand{\SiThirteen}{Si$^{13+}$}
\newcommand{\teq}{t_\mathrm{eq}}
\newcommand{\feq}{f_\mathrm{eq}}

\newcommand{\muG}{$\mu$G}
\newcommand{\TeTP}{T_{e,\mathrm{TP}}}

\newcommand{\TeNL}{T_{e,\mathrm{NL}}}
\newcommand{\TepTP}{(T_e/T_p)_\mathrm{TP}}
\newcommand{\TepNL}{(T_e/T_p)_\mathrm{NL}}
\newcommand{\xx}[1]{\!\times\!10^{#1}}
\newcommand{\RXJSNR}{SNR RX J1713.7-3946}
\newcommand{\fXRad}{f(X^i,R)}
\newcommand{\MB}{Maxwell-Boltzmann}
\newcommand{\SA}{semi-analytic}
\newcommand{\vsk}{v_\mathrm{sk}}

\shorttitle{Non-Equilibrium Ionization in Cosmic Ray Modified Shocks}
\shortauthors{Patnaude, Ellison, \& Slane}

\begin{document}

\title{The Role of Diffusive Shock Acceleration on
Non-Equilibrium Ionization in Supernova Remnants}

\author{Daniel~J. Patnaude\altaffilmark{1},
Donald~C. Ellison\altaffilmark{2}, \& Patrick Slane\altaffilmark{1}}

\altaffiltext{1}{Smithsonian Astrophysical Observatory, Cambridge, MA 02138}

\altaffiltext{2}{Physics Department, NC State University, Box 8202,
Raleigh, NC 27695; don\_ellison@ncsu.edu}

\begin{abstract}

We present results of semi-analytic calculations which show clear
evidence for changes in the non-equilibrium ionization behind a
supernova remnant forward shock undergoing efficient diffusive shock
acceleration (DSA). The efficient acceleration of particles (i.e.,
cosmic rays) lowers the shock temperature and raises the density of the
shocked gas, thus altering the ionization state of the plasma in
comparison to the test particle approximation where cosmic rays gain an
insignificant fraction of the shock energy. The differences between the
test particle and efficient acceleration cases are substantial and occur
for both slow and fast temperature equilibration rates: in cases of
higher acceleration efficiency, particular ion states are more populated
at lower electron temperatures. We also present results which show that,
in the efficient shock acceleration case, higher ionization fractions
are reached noticeably closer to the shock front than in the \TP\ case,
clearly indicating that DSA may enhance thermal X-ray production.  We
attribute this to the higher postshock densities which lead to faster
electron temperature equilibration and higher ionization rates. These
spatial differences should be resolvable with current and future X-ray
missions, and can be used as diagnostics in estimating the acceleration
efficiency in cosmic--ray modified shocks.

\end{abstract}


\section{Introduction}

In young supernova remnant (SNR) shocks, the acceleration of cosmic rays
leads to a softening of the equation of state in the shocked plasma.
This comes about because the diffusive shock acceleration (DSA)
  process turns some \nonrel\ particles into \rel\
ones and because some of the highest energy \rel\ particles escape from
the shock. Both of these effects lead to  lower
post-shock plasma temperatures as well as higher post-shock densities
\citep[e.g.,][]{JE91,BE99}. The ionization state of shocked gas at 
a particular time is
dependent upon both the gas density and the electron temperature.  In
light of this, DSA ought to leave its imprint on the ionization
structure of the shocked gas. Toward this end, we present what we
believe to be the first self-consistent model for SNR evolution which
includes the hydrodynamics, the effects of efficient shock acceleration,
and a full treatment of the non-equilibrium ionization balance at the
forward shock.

A number of young SNRs show both nonthermal and thermal emission in the
region behind the forward shock, including SN~1006
\citep{vink03,bamba08}, Tycho
\citep{hwang02,cchenai07}, and Kepler \citep{reynolds07}.  The thermal emission
arises when the forward shock sweeps up the circumstellar medium (CSM)
and heats it to X-ray emitting temperatures. As pointed out in
\citet{ellison07} (hereafter DCE07), the thermal emission is often considerably
fainter than the nonthermal emission, but there are certainly examples
where the thermal emission is as bright or brighter than any
nonthermal emission \citep{vink06}.
In \RXJSNR, the lack of thermal X-ray emission is an important
  constraint on the ambient density and significantly impacts models for
  TeV emission \citep[e.g.,][]{slane99,ESG2001,Aharonian2007,KW2008}.

If the diffusive shock acceleration process in young SNRs is as
efficient as generally believed, with $\gtrsim 50\%$ of the shock ram
kinetic energy going into \rel\ particles, nonlinear DSA will influence
the SNR hydrodynamics and be important for non-equilibrium ionization
(NEI) calculations \citep[e.g.,][]{DEB2000,EC2005}.  DCE07 took the
first steps in self-consistently coupling \NL\ DSA with NEI by tracking
the electron temperature ($T_e$) and ionization age (defined as $n_e t$,
where $n_e$ is the electron density and $t$ is the time since the material
was shocked) as a function of time in hydrodynamic simulations
of SNRs where the forward shock was efficiently producing cosmic rays
(CRs) and, as a result, was substantially modified from \TP\ results.
They found that, while both $T_e$ and $n_e t$ did differ between the
\TP\ and CR-modified cases, in the cases where DSA is highly efficient,
the \synch\ emission in the X-ray range is considerably stronger than
the thermal X-ray spectrum, and any differences in the thermal X-rays as
a result of CR-modification are likely to be missed.  In this paper, we
extend the work of DCE07 by explicitly tracking the non-equilibrium
ionization state in a CR-modified shock. The lower shock temperature and
higher density that result from efficient DSA combine to shorten both the
temperature equilibration and ionization equilibrium time-scale, 
and we show that this can have a dramatic
effect on the ionization structure between the forward shock (FS) and
the \CD\ (CD).  Although we don't calculate the thermal X-ray
emission here, the cases we study show that efficient DSA can increase
the ionization fraction of important elements and possibly enhance thermal
X-ray emission.

In \S~2, we outline the changes to our model first presented in DCE07
and discuss several caveats to our approach.  In \S~3, we present our
examples and discuss the quantitative and qualitative effects of
efficient DSA on the ionization state and SNR structure. We also show
how these effects might manifest themselves in current and future X-ray
observations. In \S~4, we summarize our results and
outline our future enhancements to this model.

\section{CR-hydro + NEI Model}

Our spherically symmetric model uses the semi-analytic DSA
  calculation developed by \citet{AB2005} and \citet{blasi05} and is
  similar to that used in DCE07, except that we now calculate the
  non-equilibrium ionization explicitly at every time step using plasma
  parameters that are continually updated as the SNR evolves.
In DCE07, the NEI was calculated at the end of the simulation using
average plasma parameters. We refer the reader to DCE07 for all details
of the CR-hydro simulation apart for those discussed below detailing our
dynamic NEI generalization.
%

The DSA model used here differs from that described in
\citet{ellison07}, \citet{EC2005}, and previous papers, in two important
ways. First, we replace the ``effective gamma,'' $\gameff$,
approximation with a more realistic model of the effect escaping
particles have on the shock dynamics. We now explicitly remove from the
shocked plasma the energy that escaping particles carry away from the
forward shock. The ratio of specific heats of the shocked gas used in
the simulations, $\gamsk$, is determined directly from the particle
distribution function including the correct mix of relativistic and
\nonrel\ particles. While the old effective gamma had the range $1 <
\gameff \le 5/3$, the ratio of specific heats $\gamsk$ is constrained to
lie between 4/3 and 5/3. These changes in the way escaping particles are
treated, and $\gameff$ is calculated, become important for later stages of
the SNR evolution, but do not produce significant changes in times as
short as 1000 yr. The results reported in \citet{ellison07} are not
modified significantly by these changes.

The second difference is that instead of specifiying a fixed injection
parameter, $\Inj$ \citep[this is $\xi$ in equation (25) in][]{blasi05},
which then determines the acceleration efficiency, we now specify a
fixed diffusive shock acceleration efficiency, $\effDSA$, and then
determine $\Inj$ accordingly. 
%
This change makes the parameterization of the acceleration efficiency
more transparent but does not change the basic approximation that is
made.

The semi-analytic DSA model we use does not calculate the acceleration
efficiency self-consistently based upon the Mach number, the available
acceleration time, and other relevant shock parameters; rather we 
parametrize the efficiency by $\Inj$, and the model then determines
the shock structure self consistently.
Furthermore, the DSA model assumes that the
thermal particles have a \MB\ distribution with a superthermal tail. The
actual shape of the quasi-thermal distribution, and the shape at the
point where the superthermal tail joins it, are approximated since the
\SA\ calculation only self-consistently describes particles with speeds
greater than the shock speed, i.e., $v_p \gg \vsk$.
The differences at low energies between what is assumed in the DSA model
and the actual quasi-thermal distribution are expected to be small, but
these differences may become more important if the contributuion to
ionization from superthermal particles is considered. Despite the
approximations of the \SA\ calculation at quasi-thermal energies, it is
the state-of-the-art since the
actual quasi-thermal distribution can only be 
determined with plasma simulations and these are not yet available for SNR
parameters.

The ionization structure of shock heated gas at a particular distance
behind the shock in a SNR is determined by the electron density $n_e$,
the electron temperature $T_e$, and the ionization and recombination
rates for each ion of interest. The structure is determined by solving
the collisional ionization equations in a Lagrangian gas element behind
the shock:

\begin{eqnarray}
\nonumber
\frac{1}{n_e} \frac{\mathrm{D}f(X^i)}{\mathrm{D}t} & = & 
C(X^{i-1},T_e)f(X^{i-1}) + \alpha(X^i,T_e)f(X^{i+1}) \\
 & & -[C(X^i,T_e) + \alpha(X^{i-1},T_e)]f(X^i) 
\ .
\end{eqnarray}

\noindent
Here, $f(X^i)$ is the fraction of element X in ion stage $X^i$ and
$C(X^i,T_e)$ and $\alpha(X^i,T_e)$ are the ionization and recombination
rates out of and into ion $X^i$, respectively.

We calculate the electron temperature by assuming that the electrons are
heated by Coulomb collisions with protons and helium
 \citep{spitzer65}. We adopt this simple prescription, which gives
   a lower limit to the equilibration time, 
knowing that the heating of electrons may, in fact, be far more
complicated. For instance, there is reason to believe that collisionless
wave-particle interactions with the magnetic turbulence will be
important \citep[e.g.,][]{laming01}, and recent work interpreting
hydrogen line widths suggests that the electron-to-proton temperature
ratio behind some SNR blast waves depends mainly on the shock speed, a
result implying a heating process substantially different from Coulomb
collisions \citep[e.g.,][]{ghavamian07,rakowski08}.  However, there
remain large uncertainties in connecting the measured line widths to the
electron-to-proton temperature ratio \citep[see][]{HS2008}, and until
particle-in-cell (PIC) simulations are able to model \nonrel,
electron-proton shocks with parameters typical of SNRs, the plasma
physics of electron heating will remain uncertain \citep[see][for a
discussion of the limitations of PIC simulations in this
regard]{VBE2008}.
In order to model some of the complexity of electron heating, we scale
the Coulomb equilibration time with a parameter, $\feq$, defined in
Eq.~(\ref{eq:eq_time}) below.

At the start of the simulation, we assume that the unshocked electrons
  and ions are in equilibrium at a temperature $T_0=10^4$\,K. We also
  assume that unshocked H and He are both 10\% singly ionized and all
  heavier elements are initially neutral. While we note that this is not
  the precise equilibrium ionization state for $10^4$\,K, we emphasize
  that none of our results depend in any significant way on the
  ionization state of the unshocked material as long as it is not fully
  neutral. In all of the results shown here we fix the helium number
  density at 10\% of the proton number density, $\np$.

At each time-step, we track the ionic state $X^i$ within each
spherically symmetric fluid element by solving the time-dependent
ionization equations for each abundant element (H, He, C, N, O, Ne, Mg,
Si, S, Ar, Ca, Fe, and Ni). We solve the coupled set of equations with
atomic data extracted from \citet{raymond77}, as first presented in
\citet{gaetz88} and updated by \citet{edgar08}.

In Figure~\ref{fig:ion_evolve} we show an example of the time evolution
of the ionization fraction, $f(X^i)$, of high ionization states of
oxygen (\OxySix, \OxySeven\ and \OxyEight) in a mass shell that is
crossed by the forward shock 100 yr after the explosion. For this
example, as in all we show in this paper, we have fixed parameters
typical of Type Ia supernovae, i.e., the kinetic energy in ejecta from
the supernova explosion $\EnSN$ = 10$^{51}$ erg, the mass of the ejecta
$\Mej$ = 1.4M$_{\sun}$, the density of the ejecta follows an
exponential density profile as is generally assumed for Type Ia
  supernovae \citep{dwarkadas00}, 
and we assume the supernova explodes in a
circumstellar medium (CSM) which is uniform with proton number density
$\np$ and magnetic field strength $B_0$.\footnote{We refer the reader to
\citet{ellison07} for a full discussion of the additional parameters
required for the CR-hydro model.}
In all of the models shown here, we take $B_0 = 15$\,\muG.\footnote{This
value for $B_0$ is somewhat higher than the typically assumed 3\,\muG\
and reflects the possibility that magnetic field amplification (MFA) may
be taking place. We emphasize, however, that we do not include MFA in
the DSA calculation performed here. A large upstream magnetic field,
$B_0$, will reduce the effects of efficient DSA, as described in
\citet{BE99}.}
The figure shows that the density and proton temperature in the
shell are dropping with time as the electron temperature increases due
to Coulomb collisions. After 1000 yr, the material is close to
ionization equilibrium for these ions.
%

Figure~\ref{fig:ion_evolve} also compares results for \TP\ (TP) and
efficient DSA.  In all of the examples in this paper, we define TP
acceleration as being 1\% efficient, i.e., 1\% of the ram kinetic energy
of the forward shock is placed into superthermal particles. For all of
our efficient acceleration cases, we assume 75\% of the shock ram
kinetic energy is placed into superthermal particles, i.e.,
$\effDSA=75\%$. Figure~\ref{fig:ion_evolve} shows that efficient DSA
produces a higher postshock density and lower postshock temperature, as
expected.  What is also clear is that the high ionization states of
oxygen become populated sooner in the $\effDSA=75\%$ case. This implies
that, instead of suppressing thermal X-ray emission as has been
suggested \cite[e.g.,][]{DAMG2008,MAB2008}, efficient DSA can possibly enhance
it.

We make the following
approximations in the NEI calculation, noting that these are in addition
to approximations made in the underlying CR-hydro model \citep[as described
in][ and references therein]{ellison07}:

\begin{itemize}

\item We assume that only electrons from the thermal population
contribute to the non-equilibrium ionization. In \NL\ DSA, the
energetic population emerges smoothly from the thermal population
\citep[a nice example from a \rel\ PIC simulation is given
in][]{Spitkovsky2008b} and superthermal particles may contribute to
ionization
\citep[see][for a \TP\ calculation involving a \MB\ distribution
  with nonthermal tail]{porquet01}. 
As we discussed above, superthermal particles are expected
  to contribute to the ionization at some level. However, the
  significance of this nonthermal ionization, in shocks undergoing
  efficient particle acceleration, has not yet been determined and
  remains an area of active work. For the purposes of this paper, we
  assume any nonthermal contribution is small.

\item We only model the interaction region between the forward shock and
the contact discontinuity where we assume cosmic elemental
abundances. One reason for emphasizing the forward shock is that it is
not certain that significant CR production occurs at the reverse shock
in SNRs \citep[e.g.,][]{EDB2005}.

\item We only consider young SNRs and do not include the effects of
radiative cooling. In the high-density limit, radiative losses could be
significant and the cooling time-scale could be comparable to other
dynamical time-scales. We will investigate this effect in a subsequent
paper.

\end{itemize}

\section{Results}
In the following examples we investigate the effect the acceleration
efficiency, $\effDSA$, and the CSM proton density, $\np$, has on the
non-equilibrium ionization state of some selected elements.

\subsection{Ionization vs. Position}

In Figure~\ref{fig:fig2}, we plot the ionization fractions of \OxySix\
and \OxySeven\,and \SiTwelve\ and \SiThirteen\, in the
top two panels as a function of position behind the forward shock
(FS). In all panels, \TP\ results ($\effDSA=1\%$) are shown with dashed
curves and efficient DSA results ($\effDSA=75\%$) are shown with solid
curves. The electron density and electron and ion temperatures are shown
in the bottom two panels.\footnote{In all results shown, we assume
that shocked protons and other ions have the same temperature.}

As the top two panels clearly show, higher ionization fractions are
attained closer to the shock front in the efficient DSA cases, as
compared to the TP cases. For instance, in the efficient case, the
fraction of \OxySeven\ peaks at a
distance $R/R_{\mathrm{FS}}\simeq 0.98$ behind the shock, while in the
TP case, this fraction peaks at $R/R_{\mathrm{FS}} \simeq 0.97$. We
attribute the increased ionization fractions closer to the shock as a
direct result of higher postshock densities in the efficient DSA case.
Note that the curves extend from the forward shock back to the contact
discontinuity, indicating that the region between the forward shock and
contact discontinuity is considerably narrower in the efficient
acceleration case. This effect produces important morphological
consequences \citep[e.g.,][]{DEB2000,WarrenEtal2005,Cassam2008}. 

In Figure~\ref{fig:fig3}, we show the same quanitities as in
Figure~\ref{fig:fig2}, except that $\np=0.1$\, \pcc. The lower CSM
density results in lower shocked densities and in less rapid collisional
ionization behind the FS. For the ions we show, higher ionization states
(i.e., \OxySeven\ and \SiTwelve) are considerably less populated
downstream from the FS when $\np$ is small. The differences resulting
from DSA are less prominent but still evident; e.g., with $\np=0.1$\,
\pcc, \OxySix\ peaks behind the shock at $R/R_{\mathrm{FS}}\simeq 0.98$
for the efficient case, and at $\simeq 0.95$ in the test particle case.



To emphasize the importance of the different spatial structures of
ionization with $\effDSA$ and $\np$, we show, in
Figure~\ref{fig:arcsec}, a closeup view of the shock fronts in
Figures~\ref{fig:fig2} and~\ref{fig:fig3}. Here, we have plotted the
ionization fractions as functions of angular distance behind the shock,
assuming a distance of 1 kpc. In the high density case ($\np=1$\,\pcc;
top panel), the fraction of \OxySix\ peaks right behind the shock at
$\sim$ 2$\arcsec$ downstream, while it peaks $\sim$ 5$\arcsec$ behind
the shock in the test particle case. 
In the lower density case ($\np=0.1$\,\pcc; lower panel), \OxySix\ peaks
$\sim$ 30$\arcsec$ behind the shock in the efficient case, but peaks
well beyond 50$\arcsec$ behind the shock in the test particle
case. Similar results are found for silicon. While these models are not
scaled to match any particular Galactic SNR, we believe the
angular separations shown here would be easily resolvable in current and
future space-based X-ray observatories even when line-of-sight
effects are taken into account. Thus, measuring the relative fraction
of H-like, He-like, and even Li-like charge
states would provide a useful diagnostic in studies of Galactic SNRs
undergoing efficient shock acceleration.

Another interesting feature seen in Figures~\ref{fig:fig2} and
\ref{fig:fig3}, is that the electron temperature is almost independent
of $\effDSA$ and only varies by a factor of $\sim 2$ between the
$\np=1$\,\pcc\ and $\np=0.1$\,\pcc\ cases. This is in contrast to the
ion temperatures, where generally lower ion temperatures occur in the
higher density models, due to the lower shock Mach number, and where the
large $\effDSA$ cases have considerably lower ion temperatures than the
\TP\ cases.
The fact that lower postshock temperatures occur in efficient DSA is
  well known \citep[e.g.,][]{Ellison2000}. The electron temperature is
  influenced by this and by the higher densities that occur with
  efficient DSA. The higher postshock densities imply more collisions
between electrons and ions, and thus more rapid temperature
  equilibration. The higher electron temperature combined with the
higher postshock density leads to more rapid ionization, and thus higher
charge states closer to the forward shock.

\subsection{Ionization vs Equilibration Timescale}
As is clear from Figures~\ref{fig:ion_evolve}, \ref{fig:fig2}, and
  \ref{fig:fig3}, the ionization fraction for high charge state ions can
  increase with acceleration efficiency.  
Since the electron temperature is almost independent of $\np$ in these
cases, we attribute this effect mainly to the higher postshock
densities.
However, we have assumed a particular model for temperature
  equilibration between protons and electrons, namely that
  electrons start off cold and equilbration with the hot protons occurs
  only through Coulomb collisions where the equilibration timescale is
  given by \citep[][Eq. 5-31]{spitzer65}:
\begin{equation}
\teq  =  \frac{3 m_p m_e k_{B}^{3/2}}{8(2\pi)^{1/2}n_p Z^2 Z_{e}^2 e^4 
\ln{\Lambda}}\left(\frac{T_p}{m_p} + \frac{T_e}{m_e}\right)^{3/2}
\ .
\label{eq:eq_time_simple}
\end{equation}
Here, $m_p$ is the proton mass and $T_p$ is the shocked proton
temperature and definitions of the other terms are given in
\citet{spitzer65}. It's important to note that
Eq.~(\ref{eq:eq_time_simple}) places strict limits on how low the
electron to proton temperature ratio can be behind the shock
\citep[see][]{HRD2000}; if other equilibration mechanisms are important,
such as plasma wave interactions, equilibration will occur more
rapidly.
To investigate the effects of more rapid
temperature equilibration, we define a parameter, $0\le \feq \le 1$, and
use the equilibration time $\teq'$ in our calculations where,
\begin{equation}
\teq'= \feq \teq
\label{eq:eq_time}
\ .
\end{equation}
In the results shown in Figures~\ref{fig:ion_evolve}, \ref{fig:fig2},
\ref{fig:fig3}, and \ref{fig:arcsec}, we have assumed $\feq=1$.

In Figure~\ref{temp_eq}, we compare the ionization fraction of
\OxySeven\ for $\effDSA=1\%$ and $\effDSA=75\%$ calculated with $\feq=1$
(black curves in all panels) and $\feq=0.1$ (red curves in all panels).
For both values of $\effDSA$, $f(O^{7+})$ is larger immediately behind
the shock for rapid equilibration ($\feq=0.1$) but drops below the
$\feq=1$ value further downstream as \OxyEight\ becomes
populated. The temperature plots in the bottom two panels show that the
electrons and protons have come into equilibrium for a range of radii
(i.e., $0.86 \lesssim R/\RadFS \lesssim 0.98$) when $\effDSA=75\%$ and
$\feq=0.1$, but remain far from equilibrium for $\feq=1$ regardless of
$\effDSA$. The equilibration rate changes the ionization structure for
this particular ion, producing changes comparible in scale to those
produced by efficient DSA.

To quantify these effects further, we look at a point midway between the
\CD\ and FS, i.e., at $\RoRfs \simeq 0.89$ for $\effDSA=75\%$ and at
$\RoRfs \simeq 0.83$ for $\effDSA=1\%$ in Figure~\ref{temp_eq}.  At
these locations, the electron to proton temperature ratios are:
$\TepTP \simeq 0.11$ and $\TepNL \simeq 0.36$, for $\feq=1$, and 
$\TepTP \simeq 0.3$ and $\TepNL = 1$ for $\feq=0.1$, i.e., the ratios
are about 3 times larger with rapid equilibration.
At these midpoint locations, the ionization fractions of \OxySeven\
range from $f($\OxySeven)\,$\simeq 0.05$ for $\feq=1$ and $\effDSA=75\%$,
to $f($\OxySeven)\,$\simeq 0.23$ for $\feq=0.1$ and $\effDSA=1\%$, i.e.,
about a factor of five span.

The electron temperature ratio for $\feq=1$ is
$(\TeNL/\TeTP)_{\feq=1}= 1.8\xx{7}\,K/2.5\xx{7}\,K \simeq 0.7$
and the ratio for $\feq=0.1$ is
$(\TeNL/\TeTP)_{\feq=0.1}= 3\xx{7}\,K/6\xx{7}\,K \simeq 0.5$. For the
particular parameters used in this example, the electron temperature
stays within a factor of $\sim 2$ for a wide spread in $\effDSA$ and
equilibration time, while $f($\OxySeven$)$ varies by a
factor of $\sim 5$.

\subsection{Emission Measure vs. Acceleration Efficiency}
%
As seen in Figures~\ref{fig:fig2} or \ref{fig:fig3}, the plasma density
is greatest immediately behind the shock where the electron temperature
is lowest. Since the rate for electron temperature equilibration depends
on the proton temperature and density and both the temperature and
density depend on $\effDSA$, the NEI calculation will depend in a
complicated fashion on the forward shock dynamics and the evolution of
the interaction region between the CD and FS. Of course, the important
property is the emission the plasma produces and this can be
characterized by the emission measure (EM) and the differential emission
measure (DEM).

In Figure~\ref{EM_plot} we plot the emission measure for individual
ions, $\mathrm{EM}=N_X \fXRad n_e n_p \mathrm{d}V$, and in
Figure~\ref{DEM_plot} we plot ionic differential emission
measures, $\mathrm{DEM} = \sum N_X \fXRad n_e n_p \mathrm{d}V/
\mathrm{d}(\log{T_e})$, where $N_X$ is the abundance of element $X$
relative to hydrogen, $\fXRad$ is the ionization fraction for the ion
$X^i$ at a distance $R$ behind the shock, and $dV$ is the volume of the
shell where EM or DEM is determined. The EM plotted in
Figure~\ref{EM_plot} is a line-of-sight projection normalized to 1
cm$^2$ surface area, and the DEM is obtained by summing over the region
between the CD and FS.

Figure~\ref{EM_plot} clearly shows that the emission for these ions
peaks much closer to the FS and is considerably stronger with efficient
DSA than in the TP case.
Figure~\ref{DEM_plot} shows that the peak emission for these two ions
shifts down in temperature by about a factor of $\sim 2$ ($\sim 1$\,keV)
when efficient DSA occurs.
These two effects are quite significant for individual ions and
  should be observable. Nevertheless, the emission from a full set of
  ions needs to be calculated and the results folded through a
  detectors' response before the signature of efficient DSA can be
  quantitatively determined.

\section{Discussion and Conclusions}
We have presented a calculation of non-equilibrium ionization in a
  hydrodynamic simulation of SNRs undergoing efficient DSA. While we
  have only explored a limited range of parameters in this paper, it's
  clear that the production of CRs by the outer blast wave modifies the
  SNR evolution and structure enough to produce significant changes in
  the ionization of the shocked material between the forward shock and
  \CD.
%
%
In particular, {\it higher ionization states are reached at lower
  electron temperatures} (compared to the test particle case) because of
  the increase in post shock density due to the increased shock
  compression.  The calculation of thermal X-ray line emission requires
  the additional step of coupling the resultant ionization state vectors
  to a plasma emissivity code, work which is in progress.  Nevertheless,
  our results clearly show that taking DSA into account and dynamically
  calculating the NEI produces changes in the ionization fractions of
  important elements that should translate into noticeable changes in
  the interpretation of X-ray line emission observed from young SNRs.

Our main results are the following:

\begin{itemize}

\item Compared to the \TP\ case, the increase in ionization that
  accompanied DSA in our examples suggests that efficient DSA
  will result
  in an increase in the overall thermal X-ray emission (see
  Figure~\ref{EM_plot}). We note that an increase in thermal
  emission with increasing acceleration efficiency is evident in our
  eariler results which explored a slightly different parameter space
\citep[i.e., Figures 7 and 8,][]{ellison07}.
The actual increase may depend importantly on other model
  parameters, such as the CSM density, and it is important to explore a
  more expanded parameter space to determine how broadly valid our results
  are. This work is in progress. However, regardless of whether or not
  efficient DSA increases the integrated thermal emission over the \TP\
  case, some thermal emission is expected because ionization is not
  suppressed when efficient DSA occurs. As Figure~\ref{fig:ion_evolve}
  shows, electrons reach X-ray emitting temperatures well before they
  come into equilibration with protons and nearly as rapidly with or
  without efficient DSA. This occurs even if only Coulomb equlibration
  is assumed. This is in contrast to recent claims
  \citep[e.g.,][]{MAB2008,DAMG2008} that very weak thermal X-ray
  emission might result from efficient shock acceleration.

\item Compared to the \TP\ case, ionization occurs more rapidly
  and, therefore, closer to the FS, with efficient acceleration (see
  Figures~\ref{fig:arcsec} and \ref{EM_plot}). The differences in
  spatial structure should be large enough to observe and may be used as
  a discriminant for the level of CR-modification, if a particular ion
  state is coupled to other known properties, such as the dynamics and
  ambient conditions.



\item Efficient DSA leads to more efficient Coulomb heating of electrons
and faster equilibration with ions, relative to the test particle
case. This results because the shocked plasma temperature is lower and
the shocked density is higher when efficient DSA occurs.
We showed, with a simple parameterization of the thermal
  equilibration time, that the signature of efficient DSA on the
  ionization state remains apparent for equilibration more rapid than
  occurs with just Coulomb collisions.

\item Using the differential
  emission measure, we showed that the maximum emission from a
  particular ion state occurs at a significantly lower electron
  temperature with efficient DSA. For the ions shown in
  Figure~\ref{DEM_plot}, the difference in $T_e$ for peak emission is on
  the order of 1 keV while the maximum DEM remains almost constant. A
  difference this large will have an important impact on the
  interpretation of thermal X-ray emission in young SNRs.


\end{itemize}

Currently, we do not treat radiative or
slow shocks, but these regimes are easily explored. For instance in a
radiative shock, the cooling time might be comparable to the energy loss
time in a cosmic ray modified shock. Increases in the density will
enhance the cooling to the point where radiative losses might rival
losses from efficient DSA \citep{wagner06}.  We intend to explore this
regime in a forthcoming paper.

While we only considered shocked CSM here, we will consider shocked
ejecta in future work. In the ejecta, the electron density can be
higher and
the temperature may be lower but, more importantly, the abundance
structure is far more complicated than for CSM and calculations of X-ray
emission are intrinsically more difficult. Furthermore, simple arguments
based on the expansion of the ejecta material suggest that the magnetic
field may be too low to support DSA by the reverse shock. Nevertheless,
there has been speculation that particles are accelerated there
\citep[e.g.,][]{GotthelfEtal2001,uchiyama08,helder08} and if DSA is
efficient at the reverse shock, it will likely alter the ionization
balance of the shocked ejecta as much as shown here for the shocked CSM.

{Finally, while we have limited our examples here to SNRs expanding into
a uniform medium typical of Type Ia supernovae, we emphasize that a wider
parameter space should be explored, in terms of both the structure of
the ambient medium (i.e. pre-SN winds) and the parameters which 
determine the cosmic ray acceleration efficiency. These cases will be 
addressed in a follow-up paper.

\acknowledgements

We would like to thank Dick Edgar, John Raymond, and Cara Rakowski 
for several useful
discussions on how to thoughtfully display and interpret ionization
fractions. This work was partially supported through a Smithsonian Endowment
Grant. D.~J.~P. and P.~O.~S. acknowledge support from NASA contract
NAS8-39073 and D.C.E achnowledges support from NASA grants
NNH04Zss001N-LTSA and 06-ATP06-21.


\begin{figure}
\epsscale{0.5}
\plotone{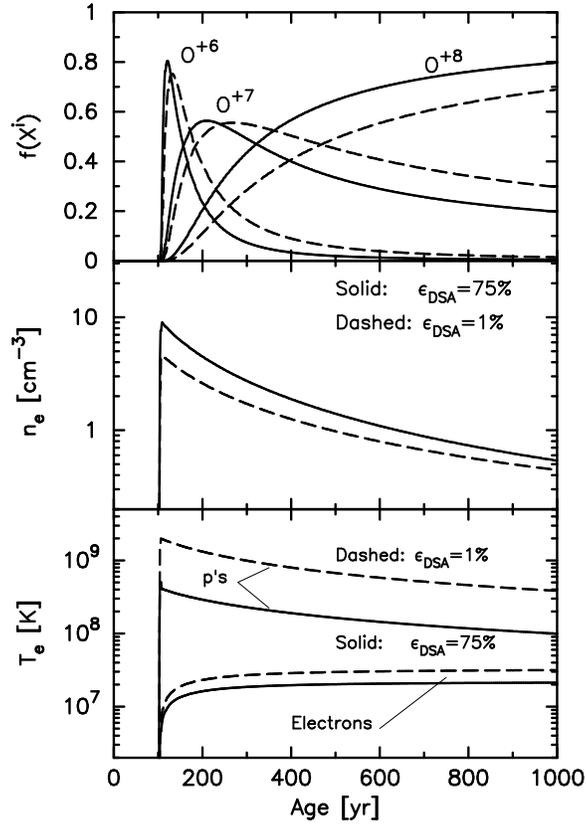}           
\caption{Time evolution of a spherically symmetric Lagrangian mass shell
which is crossed by the forward shock at 100 yr. The top panel shows the
evolution of high ionization states of oxygen, the middle panel shows
the electron number density, and the bottom panel shows the electron and
proton temperatures, assuming Coulomb equilibration. In all panels, the
solid curves correspond to a model with 75\% DSA efficiency, while the
dashed curves are for a TP model with $\effDSA=1\%$. The CSM proton
number density for this example is $\np=1$\,\pcc. Here, and in all other
examples, the unshocked CSM temperature is $T_0=10^4$\,K, and the
unshocked magnetic field is $B_0 = 15 \mu\mathrm{G}$.}
\label{fig:ion_evolve}
\end{figure}

\begin{figure}
\epsscale{0.5}
\plotone{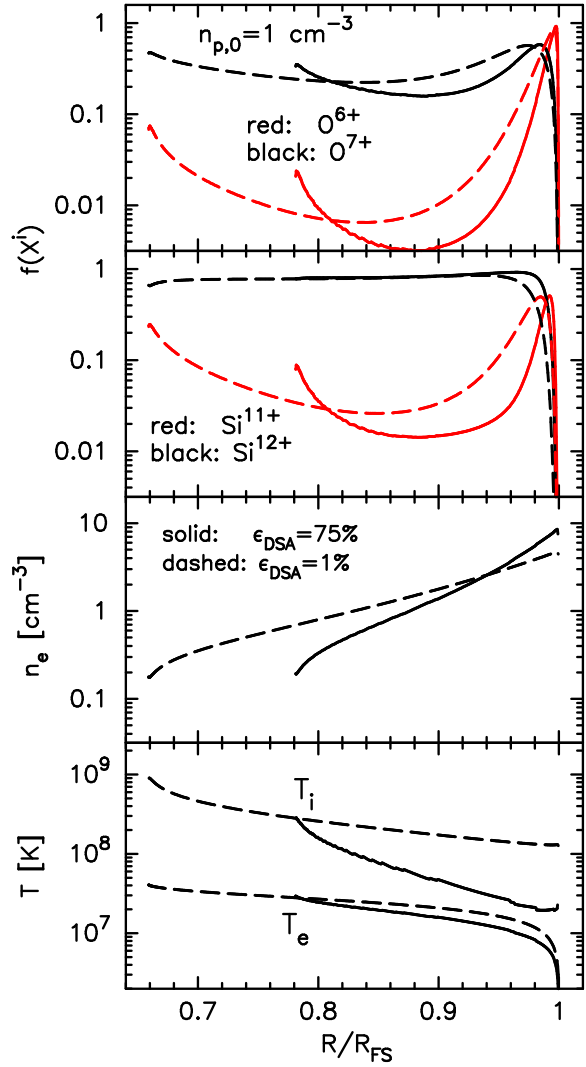}              
\caption{Spatial profiles of H- and He-like oxygen and silicon, electron
density, and temperature as a function of distance behind the forward
shock. In the bottom panel, the curves labeled $T_i$ are ion (or
  proton) temperatures and those $T_e$ are electron temperatures.
Here, and in 
figures~\ref{fig:fig3}-\ref{temp_eq} that follow, we show values from
spherically symmetric shells as a function of $R$ or $\Delta R$, 
not line-of-sight projections.  In all panels, solid curves correspond
to models with 75\% efficiency, while the dashed lines correspond to TP
models.  These models are for a CSM proton density of $\np = 1$
cm$^{-3}$ and are calculated at $\tSNR=1000$\,yr. In the model with
75\% efficiency, the forward shock velocity is $\approx$ 1800 km s$^{-1}$, 
while in the test particle model, it is $\approx$ 2200 km s$^{-1}$
at $\tSNR=1000$\,yr.}
\label{fig:fig2}
\end{figure}

\begin{figure}
\epsscale{0.5}
\plotone{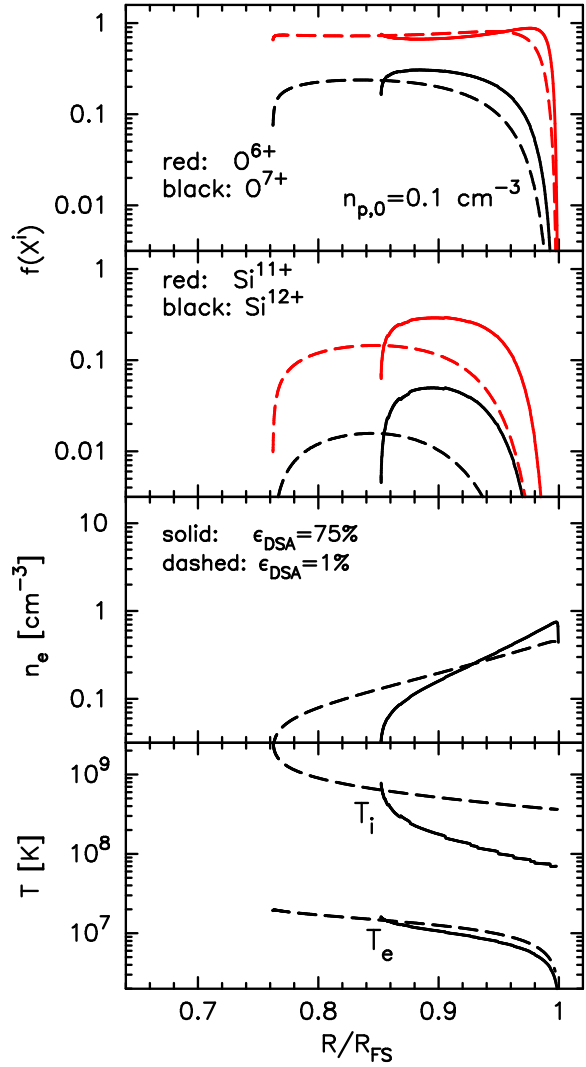}     
\caption{Spatial profiles of oxygen and
silicon ions, electron density, and temperature as a function of
distance behind the forward shock. In all panels, solid curves
correspond to models with 75\% efficiency, while the dashed lines
correspond to TP models.  These models are for a CSM proton density $\np
= 0.1$ cm$^{-3}$ and are calculated at $\tSNR=1000$\,yr. In the model with
75\% efficiency, the forward shock velocity is $\approx$ 3200 km s$^{-1}$, 
while in the test particle model, it is $\approx$ 3600 km s$^{-1}$
at $\tSNR=1000$\,yr.}
\label{fig:fig3}
\end{figure}

\begin{figure}
\epsscale{0.5} \plotone{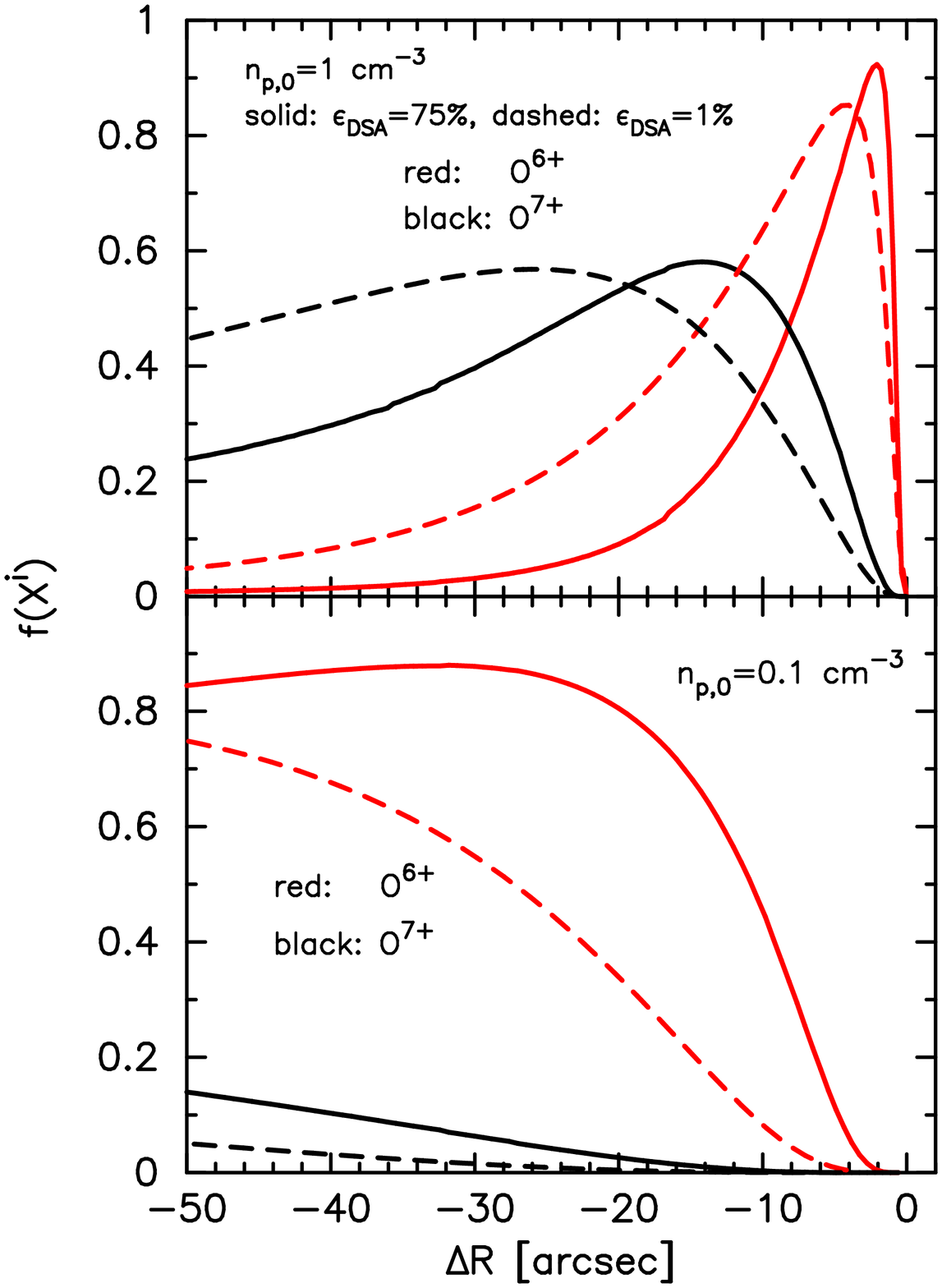}   
\caption{{\it Top:} Ionization fraction as a
function of distance behind the forward shock for \OxySix\ and
\OxySeven\ with $\np = 1.0$\,\pcc.  {\it Bottom:} Ionization fractions
of \OxySix\ and \OxySeven\ with $\np=0.1$\,\pcc. In both panels, the
solid curves are for $\effDSA=75\%$ and the dashed curves are for
$\effDSA=1\%$. The angular scale is
determined assuming the SNR is at a distance of 1 kpc and
the results are calculated at $\tSNR=1000$\,yr.}
\label{fig:arcsec}
\end{figure}

\begin{figure}
\epsscale{0.5} \plotone{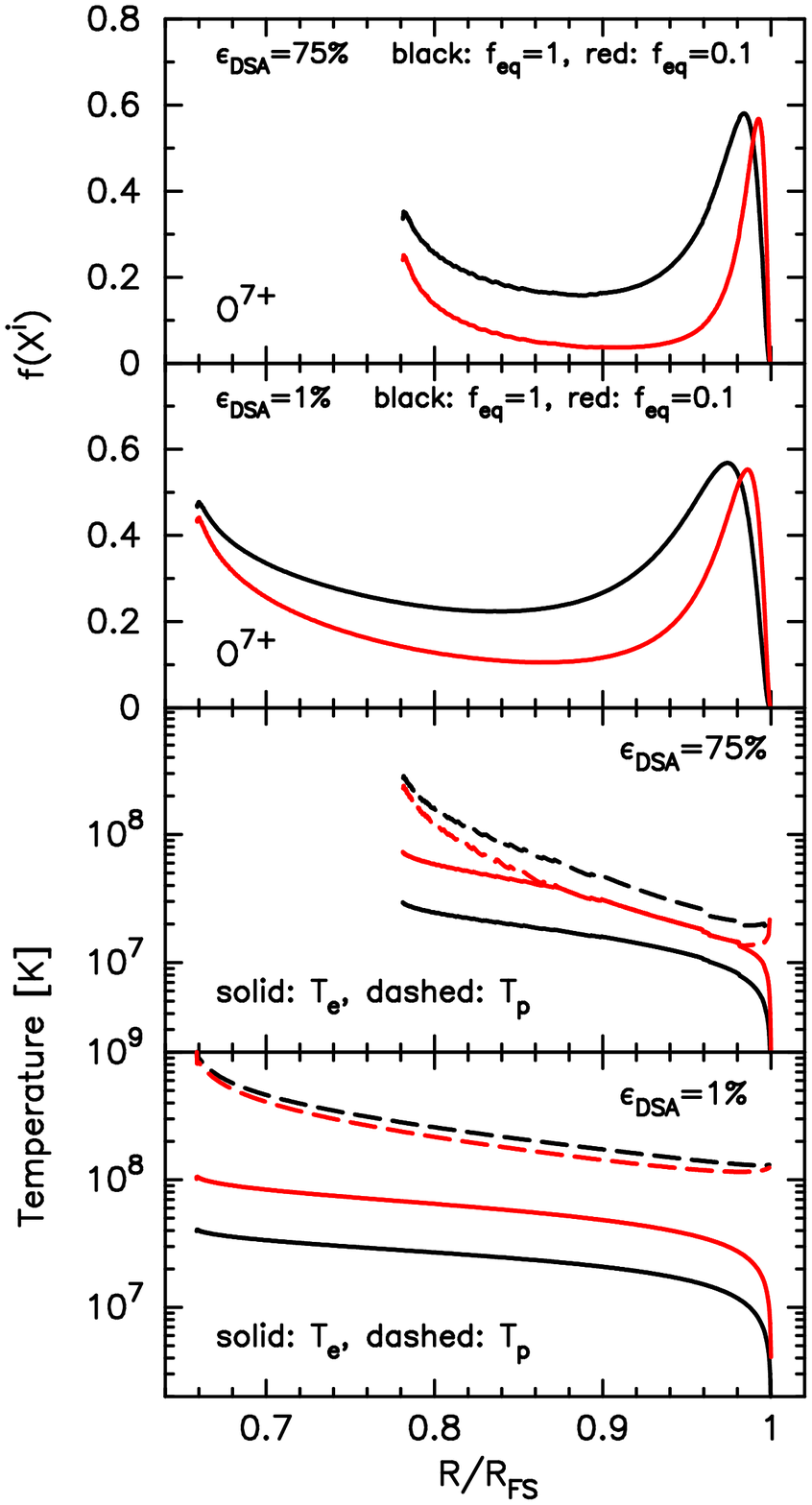}  
\caption{Ionization fraction and temperature
calculated between the \CD\ and FS. All calculations are at
$\tSNR=1000$\,yr and assume $\np=1$\,\pcc. In all panels, black curves
assume $\feq=1$ and red curves assume $\feq=0.1$. In the bottom two
panels, the solid curves are the shocked electron temperature, $T_e$,
and the dashed cuvres are the shocked proton temperature, $T_p$. As in
Figures~\ref{fig:fig2} and \ref{fig:fig3}, the left end of each curve is
at the position of the CD.
\label{temp_eq}}
\end{figure}

\begin{figure}
\plotone{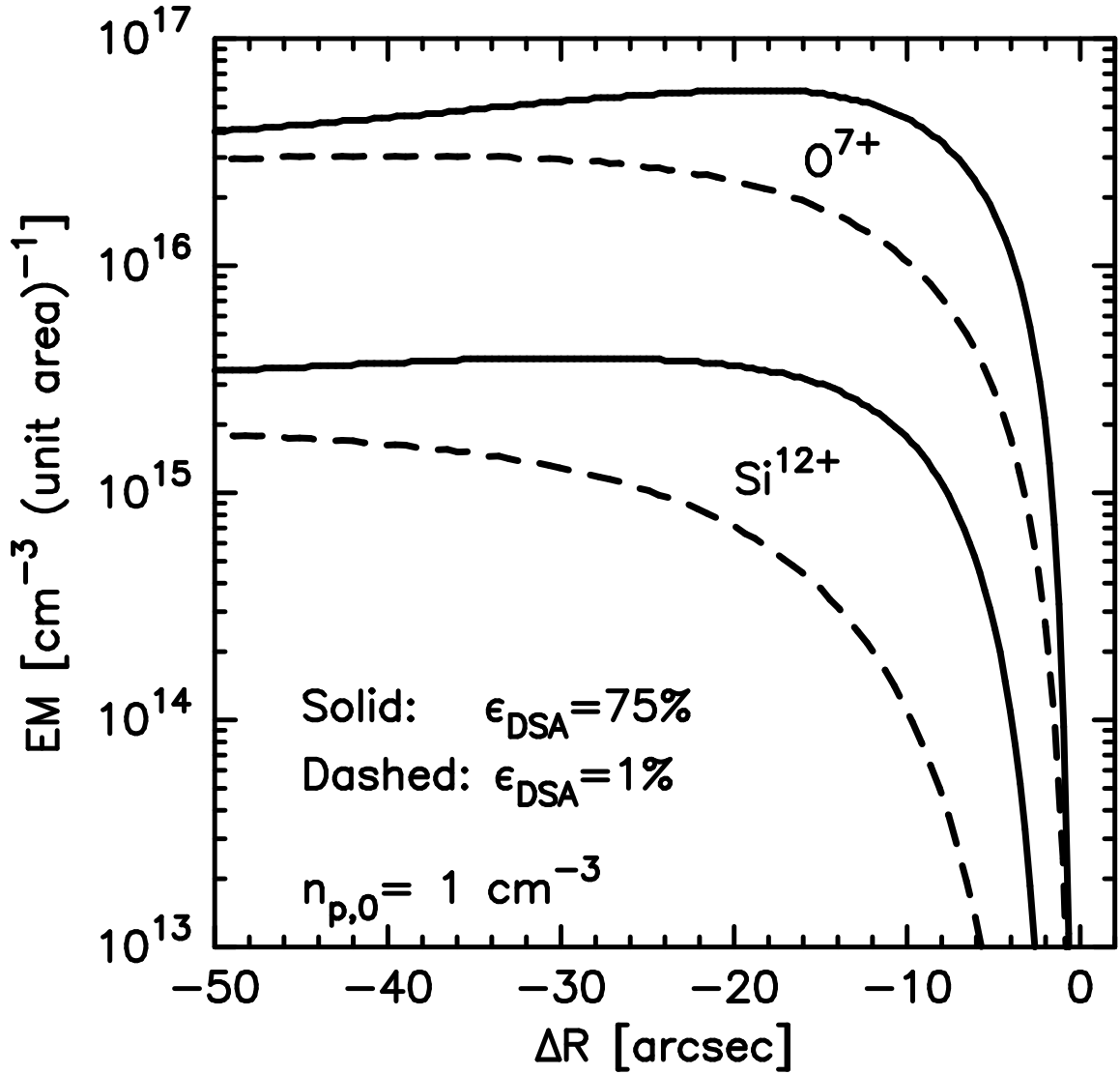}  
\caption{Line-of-sight projection of the emission measure (EM) for
  \OxySeven\ and \SiTwelve\ as labeled. The solid curves are for
  $\effDSA=75\%$ and the dashed curves are for $\effDSA=1\%$. The
  angular distance, $\Delta R$, from the FS is determined assuming the
  SNR is at 1 kpc and the results are calculated at $\tSNR=1000$\,yr
  with $\np=1$\,\pcc\ and $\teq=1$.
\label{EM_plot}}
\end{figure}

\begin{figure}
\plotone{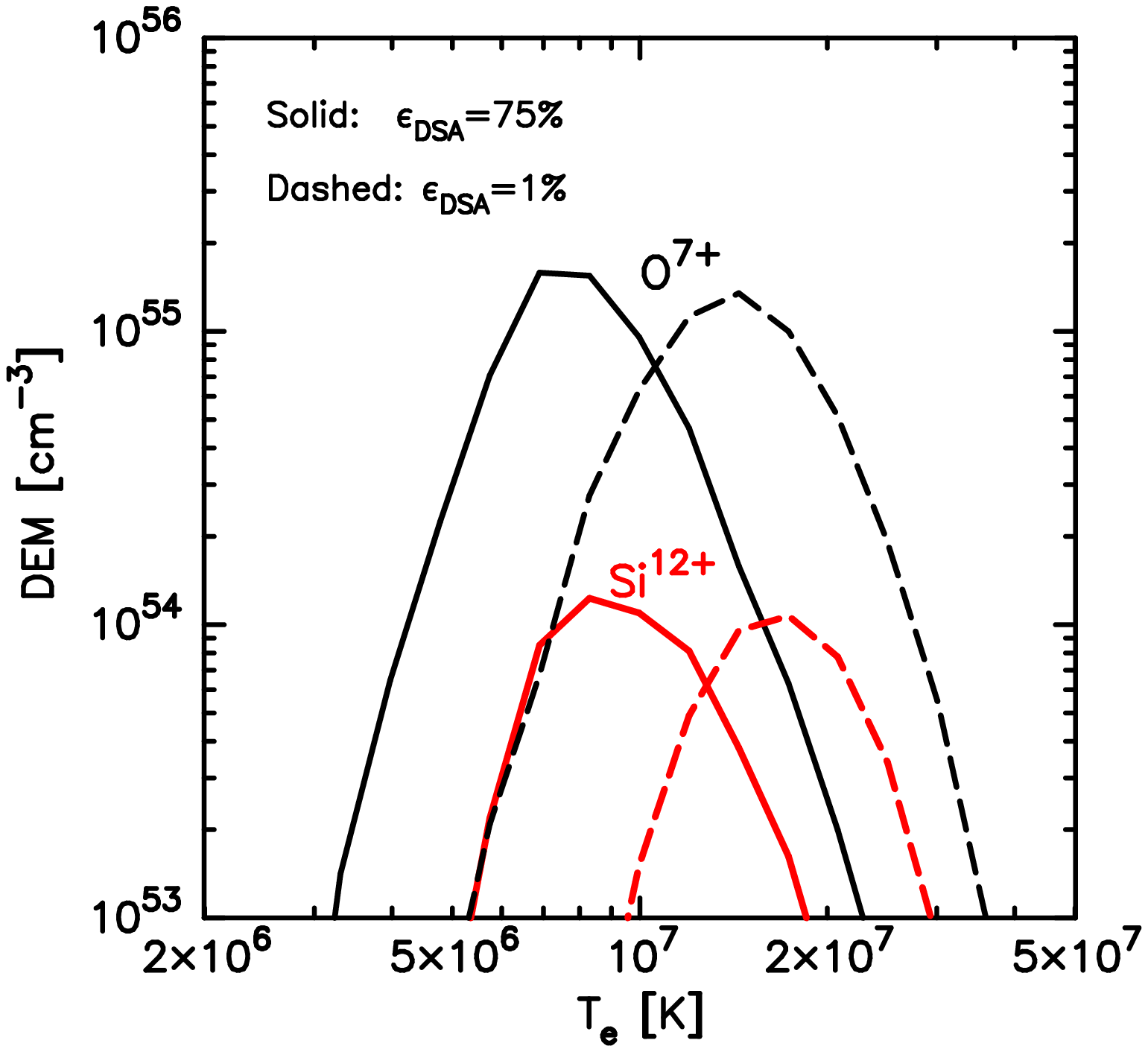}  
\caption{Differential emission measure (DEM) vs. electron temperature
  for \OxySeven\ and \SiTwelve\ as labeled. The solid curves are for
  $\effDSA=75\%$ and the dashed curves are for $\effDSA=1\%$. The
  results are calculated at $\tSNR=1000$\,yr with $\np=1$\,\pcc\
  and $\teq=1$.
\label{DEM_plot}}
\end{figure}

\end{document}